\documentclass[twocolumn,showpacs,superscriptaddress,preprintnumbers,amsmath,amssymb,prc]{revtex4-2}

\usepackage{graphicx}% Include figure files
\usepackage{dcolumn}% Align table columns on decimal point
\usepackage{bm}% bold math
\usepackage{float}
\usepackage{color}
\usepackage{CJK}
\usepackage{ulem}
\usepackage{array} 
\usepackage{makecell}
\usepackage{multirow}

\usepackage[colorlinks,linkcolor=blue,urlcolor=blue,citecolor=blue]{hyperref}

\usepackage{tikz,xcolor,hyperref}
\definecolor{lime}{HTML}{A6CE39}

\DeclareRobustCommand{\orcidicon}{
	\begin{tikzpicture}
	\draw[lime, fill=lime] (0,0) 
	circle [radius=0.16] 
	node[white] {{\fontfamily{qag}\selectfont \tiny ID}};
	\draw[white, fill=white] (-0.0625,0.095) 
	circle [radius=0.007];
	\end{tikzpicture}
	\hspace{-2mm}
}
\foreach \x in {A, ..., Z}{%
	\expandafter\xdef\csname orcid\x\endcsname{\noexpand\href{https://orcid.org/\csname orcidauthor\x\endcsname}{\noexpand\orcidicon}}
}

\begin{document}
%\begin{CJK*}{GBK}{}
\begin{CJK*} {UTF8} {gbsn}

\title{Simulations of momentum correlation functions of light (anti)nuclei in relativistic heavy-ion collisions at $\sqrt{s_{NN}}$ = 39 GeV}

\author{Ting-Ting Wang(王婷婷)}
\affiliation{Key Laboratory of Nuclear Physics and Ion-Beam Application (MOE), Institute of Modern Physics, Fudan University, Shanghai 200433, China}

\author{Yu-Gang Ma(马余刚)\orcidB{}} \thanks{Corresponding author:  mayugang@fudan.edu.cn}
%\footnote{Corresponding author: mayugang@fudan.edu.cn}}
\affiliation{Key Laboratory of Nuclear Physics and Ion-Beam Application (MOE), Institute of Modern Physics, Fudan University, Shanghai 200433, China}
\affiliation{Shanghai Research Center for Theoretical Nuclear Physics， NSFC and Fudan University, Shanghai 200438, China}

\author{Song Zhang(张松)\orcidC{} %\orcid{https://orcid.org/0000-0003-2782-7801}
}
%\thanks{Email: song\_zhang@fudan.edu.cn}
\affiliation{Key Laboratory of Nuclear Physics and Ion-Beam Application (MOE), Institute of Modern Physics, Fudan University, Shanghai 200433, China}
\affiliation{Shanghai Research Center for Theoretical Nuclear Physics， NSFC and Fudan University, Shanghai 200438, China}

\date{\today}

\begin{abstract}
Momentum correlation functions of light (anti)nuclei formed by the coalescence mechanism of (anti)nucleons are calculated for several central  heavy-ion collision systems, namely $_{5}^{10}\textrm{B}+_{5}^{10}\textrm{B}$, $_{8}^{16}\textrm{O}+_{8}^{16}\textrm{O}$, $_{20}^{40}\textrm{Ca}+_{20}^{40}\textrm{Ca}$ as well as $_{79}^{197}\textrm{Au}+_{79}^{197}\textrm{Au}$ in different centralities at center of mass energy $\sqrt{s_{NN}}$ = 39 GeV within the framework of  A Multi-Phase Transport (AMPT) model complemented by the Lednick$\acute{y}$ and Lyuboshitz analytical method. 
Momentum correlation functions for identical or nonidentical light (anti)nuclei are %for the first time 
constructed and analyzed for the above collision systems.  The  Au + Au  results demonstrate that emission of light (anti)nuclei  occurs from a source with smaller space extent in more peripheral collisions. The effect of system-size on the momentum correlation functions of identical or nonidentical light (anti)nuclei is also explored by several collision system in central collisions. 
The results indicate that the emission source-size of light (anti)nuclei pairs deduced from their momentum correlation functions and system-size is self-consistent. Momentum correlation functions of nonidentical light nuclei pairs gated on velocity are applied to infer the average emission sequence of them. The results illustrate that protons are emitted in average on a similar time scale with neutrons but earlier than deuterons or tritons in the small relative momentum region. In addition, larger interval of the average emission order among them is exhibited for  smaller collision systems or at more peripheral collisions.

\end{abstract}

%\pacs{25.70.Mn, 24.10.-i, 25.70.Pq, 27.80.+w}

\maketitle
\section{Introduction}
In heavy-ion collisions (HICs), two-particle momentum correlation function is different from the original application in astronomy~\cite{Hanbury1956-1,Hanbury1956-2}, and has been normally utilized to extract space-time information of the emission source and probe the dynamical evolution of nuclear collisions in an extensive energy range~\cite{Koonin1977,Lisa2005,Heinz,Goldhaber1960,Bialas2000,Ghetti2000,Boal1990,Ardouin1997,RGhetti2003,HeJJ}. Many different studies on the two-particle momentum correlation functions in intermediate energy HICs can be also found in literature, eg. Refs.~\cite{wtt2018,Kotte2005,RGhetti2003,DGourio2000,Poch1987,WGGong1991,YGMa2006,RGhetti2004,LWChen2003,DQFang2016,HuangBS2,ShenLei,WangTT_2022,Fang,Li1_SCPMA,Li2_SCPMA}, which include the momentum correlation functions of neutron, proton as well as light charged particle (LCP) pairs. Multi-variable dependences of the momentum correlation functions, such as impact parameters, total momentum of particle pairs, isospin of the emission source, nuclear symmetry energy, nuclear equation of state (EOS) as well as in-medium nucleon-nucleon cross section (NNCS) etc., contain a wealth of information about the space-time characteristics of intermediate energy HICs. In high energy HICs,  two-hadron momentum correlation function， also called as Hanbury Brown-Twiss (HBT) interferometry, 
was also well extensively  measured and some interesting properties on emission source were extracted 
\cite{STAR1,STAR2}.
 Oscillations of the extracted HBT radii versus emission angle indicate that emission  source is elongated perpendicular to the reaction plane.   The results indicate that the initial shape is more or less remained and could be identified even though the collision system undergoes the pressure and expansion.
Furthermore, interaction between antiprotons has been also measured with the momentum correlation functions and the equality of interactions between $p$-$p$ and $\bar{p}$-$\bar{p}$ was  proved by the STAR Collaboration~\cite{Star-nature}. The interaction property of the particle pairs has been discussed for other particles, for instance $\Lambda$ pairs~\cite{Star-prl}, proton-$\Omega$ and proton-$\Xi$ etc \cite{Star-pOmega,Alice-Nature}, with the same momentum correlation technique. Furthermore, the measurements of  momentum correlation functions for nonidentical nucleons and light clusters can be used to characterize the mean emission sequence of them, which was firstly proposed in Ref.~\cite{Gelderloos1994}. Theoretical study has been extended to different kinds of nonidentical particle pairs, for instance $p$-$d$, $n$-$p$~\cite{Lednicky1996,wtt2019,YijieWang2022,XiBS}, $\pi$-$p$~\cite{Voloshin1997}, $K^{+}$-$K^{-}$~\cite{Ardouin1999}, $d$-$t$~\cite{DGourio2000,RGhetti2003} as well as $^{3}He$-$\alpha$ particles~\cite{Kotte1999} in intermediate energy HICs.

In this work we extend the studies, for the first time, on the momentum correlation functions of light (anti)nuclei to ultra-relativistic heavy-ion collisions  simulated by  A Multi-Phase Transport (AMPT) model~\cite{ZWLin2005,ZWLin-new} coupled with a dynamical coalescence model~\cite{Chen2003npa,SZhang2010,YiLinCheng2021}, specifically at $\sqrt{s_{NN}}$ = 39 GeV. Different gating conditions such as centrality gate, system-size gate as well as velocity gate are applied to the momentum correlation functions of light (anti)nuclei pairs. In particular, we report on the indication of  the emission chronology of protons, deuterons and tritons which can be deduced from their corresponding momentum correlation functions in ultra-relativistic  HICs at $\sqrt{s_{NN}}$ = 39 GeV. The emission sequence~of light clusters inferred from the correlation functions is expected measurable in future experiments to verify our deduction from the coalescence picture.

The rest of this article is organized as follows. In Section II A and II B, we briefly describe A Multi-Phase Transport  model~\cite{ZWLin2005,ZWLin-new} and the coalescence model~\cite{Chen2003npa,SZhang2010,YiLinCheng2021}, then introduce how to calculate the momentum correlation functions of particle pairs by using the Lednick$\acute{y}$ and Lyuboshitz analytical formalism~\cite{Koonin1977,Lednicky2007,lednicky2006,lednicky2009,lednicky2008} in Section II C. In Section III, we summarize the simulated results of the light (anti)nuclei momentum correlation functions gated on various parameters in relativistic heavy-ion collisions. Section III A compares the results of proton-proton and proton-antiproton momentum correlation functions with experimental data from the RHIC-STAR collaboration. From Section III B to III D, identical and nonidentical light (anti)nuclei momentum correlation functions gated on different conditions are systematically discussed. Finally, a summary and outlook are given in Section IV.

\section{MODELS AND FORMALISM}
\subsection{AMPT model}

To obtain phase-space distributions of (anti)particles, A Multi-Phase Transport  model~\cite{ZWLin2005,ZWLin-new} is used as the event generator, which has been applied successfully for studying heavy-ion collisions at relativistic energies, eg. ~\cite{SZhang2010,Alver2010,GLMa1,LXHan2011,GLMa2,zzq2014,SZhang2017,YiLinCheng2021,ZhangH2021,WangH21,WangH22}. We briefly review the main components of the AMPT model used in the present work. In the version of AMPT, the initial phase-space information of partons is generated by the heavy-ion jet interaction generator (HIJING) model~\cite{XNWang1991,MGyulassy1994}. The interaction between partons is then simulated by Zhang's parton cascade (ZPC) model~\cite{BZhang1998}. During the hadronization process, a quark coalescence model is used to combine partons into hadrons~\cite{ZWLin2001,Subrata2004,Subrata2002}. Then, the hadronic rescattering evolution is described by a relativistic transport (ART) model~\cite{BALi1995}. 

In this paper, the collisions of $_{5}^{10}\textrm{B}+_{5}^{10}\textrm{B}$, $_{8}^{16}\textrm{O}+_{8}^{16}\textrm{O}$, $_{20}^{40}\textrm{Ca}+_{20}^{40}\textrm{Ca}$ at $0-10$ $\%$ centrality and  mid-rapidity ($\left|y \right|< 0.5$) as well as  $_{79}^{197}\textrm{Au}+_{79}^{197}\textrm{Au}$ at same mid-rapidity for five centralities of $0-10$ $\%$, $10-20$ $\%$, $20-40$ $\%$, $40-60$ $\%$, and $60-80$ $\%$ at $\sqrt{s_{NN}}$ = 39 GeV are simulated. 
Te phase-space distributions of (anti)particles are  selected at the final stage in the hadronic rescattering process (ART model~\cite{BALi1995}) with considering baryon-baryon, baryon-meson, and meson-meson elastic and inelastic scatterings, as well as resonance decay or week decay. The transverse momentum spectra of light (anti)nuclei  have been successfully reproduced by the AMPT model with the maximum hadronic rescattering time (MRT) of 100 fm/c~\cite{YiLinCheng2021}. Therefore, the same maximum hadronic rescattering time is used for the most  calculations in this work except for a quantitative comparison with the $p$-$p$ and $p$-$\bar{p}$ data from the STAR collaboration in Sec. III A.

\subsection{Coalescence model}

The coalescence model has been used widely in describing the production of light clusters in the intermediate~\cite{LieWenChen2003prc,MGyulassy1983,JAichelin1987,VolkerKoch1990,Pawlowski2000} and  high~\cite{Mattiello1997,Nagle1996} energy heavy-ion collisions.~The detailed definitions of the probability for producing a cluster of nucleons is  in Ref.~\cite{Chen2003npa}. In our model calculations, light (anti)clusters such as (anti)deuterons and tritons are constructed by using the coalescence model as follows~\cite{Csernai1986,ChenJH2018}.
 The probability for producing $M$-nucleon cluster is determined by its Wigner phase-space density and the nucleon phase-space distribution at the freeze-out stage~\cite{Chen2003npa}. The multiplicity of an $M$-nucleon cluster in transport model simulations for heavy-ion collisions is given by,

\begin{multline}
N_{M}=G \int \sum_{i_{1}>i_{2}>\cdots>i_{M}} d \vec{r}_{i_{1}} d \vec{k}_{i_{1}} \cdots d \vec{r}_{i_{M-1}} d \vec{k}_{i_{M-1}} \\
\left\langle\rho_{i}^{W}\left(\vec{r}_{i_{1}}, \vec{k}_{i_{1}}, \cdots, \vec{r}_{i_{M-1}}, \vec{k}_{i_{M-1}}\right)\right\rangle 
\end{multline}
\hspace*{\fill} \\
where $\vec{r}_{i_{1}}, \vec{r}_{i_{M-1}}$ and $\vec{k}_{i_{1}}, \vec{k}_{i_{M-1}}$ are the relative coordinates and momentum in the $M$-nucleon rest frame, and spin-isospin statistical factor $G$ is 3/8 for (anti)deuteron and 1/3 for triton~\cite{Chen2003npa}. In addition,  $\rho^{W}$ is the Wigner density function, which is different for all kinds of particles. Therefore, we will calculate separately the Wigner phase-space density of (anti)deuteron and triton in detail. The Wigner phase-space density of (anti)deuteron is constructed by,

\begin{multline}
\rho_{d}^{W}(\vec{r}, \vec{k})=8 \sum_{i=1}^{15} c_{i}^{2} \exp \left(-2 \omega_{i} r^{2}-\frac{k^{2}}{2 \omega_{i}}\right) \\
+16 \sum_{i>j}^{15} c_{i} c_{j}\left(\frac{4 \omega_{i} \omega_{j}}{\left(\omega_{i}+\omega_{j}\right)^{2}}\right)^{\frac{3}{4}} \exp \left(-\frac{4 \omega_{i} \omega_{j}}{\omega_{i}+\omega_{j}} r^{2}\right) \\
\times \exp \left(-\frac{k^{2}}{\omega_{i}+\omega_{j}}\right) \cos \left(2 \frac{\omega_{i}-\omega_{j}}{\omega_{i}+\omega_{j}} \vec{r} \cdot \vec{k}\right)
\end{multline}
\hspace*{\fill} \\
where $\vec{k}=\left(\vec{k}_{1}-\vec{k}_{2}\right) / 2$ is the relative momentum and $\vec{r}=\left(\vec{r}_{1}-\vec{r}_{2}\right)$ is the relative coordinate of (anti)proton and (anti)neutron.  The Wigner phase-space density of triton is constructed by a spherical harmonic oscillator~\cite{SZhang2010,Chen2003npa,KJSun2015},

\begin{multline}
\rho_{t}^{W}\left(\rho, \lambda, \vec{k}_{\rho}, \vec{k}_{\lambda}\right)\\
=\int \psi\left(\rho+\frac{\vec{R}_{1}}{2}, \lambda+\frac{\vec{R}_{2}}{2}\right) \psi^{*}\left(\rho-\frac{\vec{R}_{1}}{2}, \lambda-\frac{\vec{R}_{2}}{2}\right) \\
 \times \exp \left(-i \vec{k}_{\rho} \cdot \vec{R}_{1}\right) \exp \left(-i \vec{k}_{\lambda} \cdot \vec{R}_{2}\right) 3^{\frac{3}{2}} d \vec{R}_{1} d \vec{R}_{2} \\
=8^{2} \exp \left(-\frac{\rho^{2}+\lambda^{2}}{b^{2}}\right) \exp \left(-\left(\vec{k}_{\rho}^{2}+\vec{k}_{\lambda}^{2}\right) b^{2}\right)
\end{multline}
\hspace*{\fill} \\
where $\rho$ and $\lambda$ are relative coordinates, $\vec{k}_{\rho}$ and $\vec{k}_{\lambda}$ are the relative momenta in the Jacobi coordinate.

The above parameters of the Gaussian fit coefficient $c_{i}$ and $w_{i}$ for (anti)deuteron as well as $b$ for triton are given in Ref.~\cite{Chen2003npa}. Based on the phase-space information of light (anti)cluster obtained by the above coalescence model, the momentum correlation functions of (non)identical light (anti)cluster pairs can be discussed in the following.

\subsection{Lednik\texorpdfstring{$\acute{y}$}  and Lyuboshitz technique}

Next, we briefly review the technique of the two-particle momentum correlation function proposed by Lednick$\acute{y}$ and Lyuboshitz~\cite{lednicky2006,lednicky2009,lednicky2008}. The method  is based on the principle as follows: when two particles are emitted at small relative momentum, their momentum correlation function is determined by the space-time characteristics of the production processes owing to the effects of quantum statistics (QS) and final-state interactions (FSI)~\cite{Koonin1977,Lednicky2007}. The details on the formalism of the two-particle momentum correlation function can be found in Ref.~\cite{wtt2019}.

Here, comparing with our previous literature~\cite{wtt2019}, more particle pairs are considered in the article. Therefore, the final-state interaction of different particle pairs can be known well by introducing $f_c\left(k^*\right)$ particularly as follows:
\begin{equation}
f_c\left(k^*\right) = \left[ K_{c}\left(k^*\right)-\frac{2}{a_c}h\left(\lambda \right)-ik^*A_{c}\left(\lambda \right)\right]^{-1}
\end{equation}
\hspace*{\fill} \\
$f_c\left(k^*\right)$ is the $s$-wave scattering amplitude renormalizied by the long-range Coulomb interaction, with $h\left(\lambda \right) = \lambda^{2}\sum_{n=1}^{\infty}\left[n\left(n^2+\lambda^2\right)\right]^{-1}-C-\ln\left[\lambda \right]$ where $C$ = 0.5772 is the Euler constant. $K_{c}\left(k^*\right) = \frac{1}{f_0} + \frac{1}{2}d_0k^{*^2} + Pk^{*^4} + \cdots$ is the effective range function, where $d_{0}$ is the effective radius of the strong interaction, $f_{0}$ is the scattering length and $P$ is the shape parameter. The parameters of  effective range function are important to characterize the essential properties of the final-state interactions, and can be extracted from the correlation function measured experimentally~\cite{Erazmus1994,Arvieux1974,Star-nature,wtt2019}. Table I shows the parameters of the effective range function for different particle pairs in the present work.

\begin{table}[!htbp]
\caption{Experimental determination of the effective range function parameters for $n$-$n$ ($\bar{n}$-$\bar{n}$), $p$-$p$ ($\bar{p}$-$\bar{p}$), $t$-$t$, $n$-$p$ ($\bar{n}$-$\bar{p}$), $p$-$d$ ($\bar{p}$-$\bar{d}$), $p$-$t$ and $d$-$t$ systems \cite{Erazmus1994,Star-nature,Arvieux1974}.}
\hspace*{\fill} \\
\begin{tabular}{ccccc}
\hline
\hline
\multicolumn{1}{c}{System} & Spin & $f_{0}$ $\left(fm\right)$ & $d_{0}$ $\left(fm\right)$ & $P$ $\left(fm^{3}\right)$ \\ \hline
$n$-$n$ ($\bar{n}$-$\bar{n}$)                          & 0    & 17     & 2.7     & 0.0     \\
$p$-$p$ ($\bar{p}$-$\bar{p}$)                          & 0    & 7.8     & 2.77     & 0.0     \\
$t$-$t$                          & 0    & $1\times10^{-6}$  & 0.0      & 0.0     \\
$n$-$p$ ($\bar{n}$-$\bar{p}$)                          & 0    & 23.7  & 2.7      & 0.0      \\
\multirow{2}{*}{$p$-$d$ ($\bar{p}$-$\bar{d}$)}         & 1/2  & -2.73   & 2.27     & 0.08    \\
                            & 3/2  & -11.88  & 2.63     & -0.54   \\
$p$-$t$                          & 0    & $1\times10^{-6}$  & 0.0      & 0.0     \\
$d$-$t$                          & 0    & $1\times10^{-6}$  & 0.0      & 0.0     \\ \hline
\hline
\end{tabular}
\end{table}
\hspace*{\fill} 

In the table I, for $n$-$n$ ($\bar{n}$-$\bar{n}$) and $n$-$p$ ($\bar{n}$-$\bar{p}$) momentum correlation functions which include uncharged particle, the Coulomb penetration factor ($A_c\left(\lambda \right)$) is not considered and only the short-range particle interaction works. For the momentum correlation functions of charged particles such as $p$-$\bar{p}$, $p$-$p$ ($\bar{p}$-$\bar{p}$), $d$-$d$ ($\bar{d}$-$\bar{d}$), $t$-$t$, $p$-$d$ ($\bar{p}$-$\bar{d}$), $p$-$t$ and $d$-$t$, both the Coulomb interaction and the short-range interaction dominated by the $s$-wave interaction are taken into account. The momentum correlation function of $p$-$p$ ($\bar{p}$-$\bar{p}$) particle pairs is dominantly contributed by only the singlet $\left(S = 0\right)$ $s$-wave  final-state interactions while both spins 1/2 and 3/2 contribute in the case of $p$-$d$ ($\bar{p}$-$\bar{d}$) system. Moreover, for (anti)deuteron-(anti)deuteron momentum correlation function, a parametrization of the $s$-wave phase shifts $\delta$ has been used from the solution of $K_{c}\left(k^*\right) = \cot{\delta}$ for each total pair spin $S = 0, 1, 2$. Note that the effective range function for the total spin $S = 1$ is irrelevant, since it does not contribute due to the quantum statistics symmetrization.

\section{ANALYSIS AND DISCUSSION}

\subsection{Comparison between our $p$-$p$ and  $p$-$\bar{p}$ correlation functions with experimental data}

\begin{figure}[!htbp]
 \includegraphics[width=\linewidth]{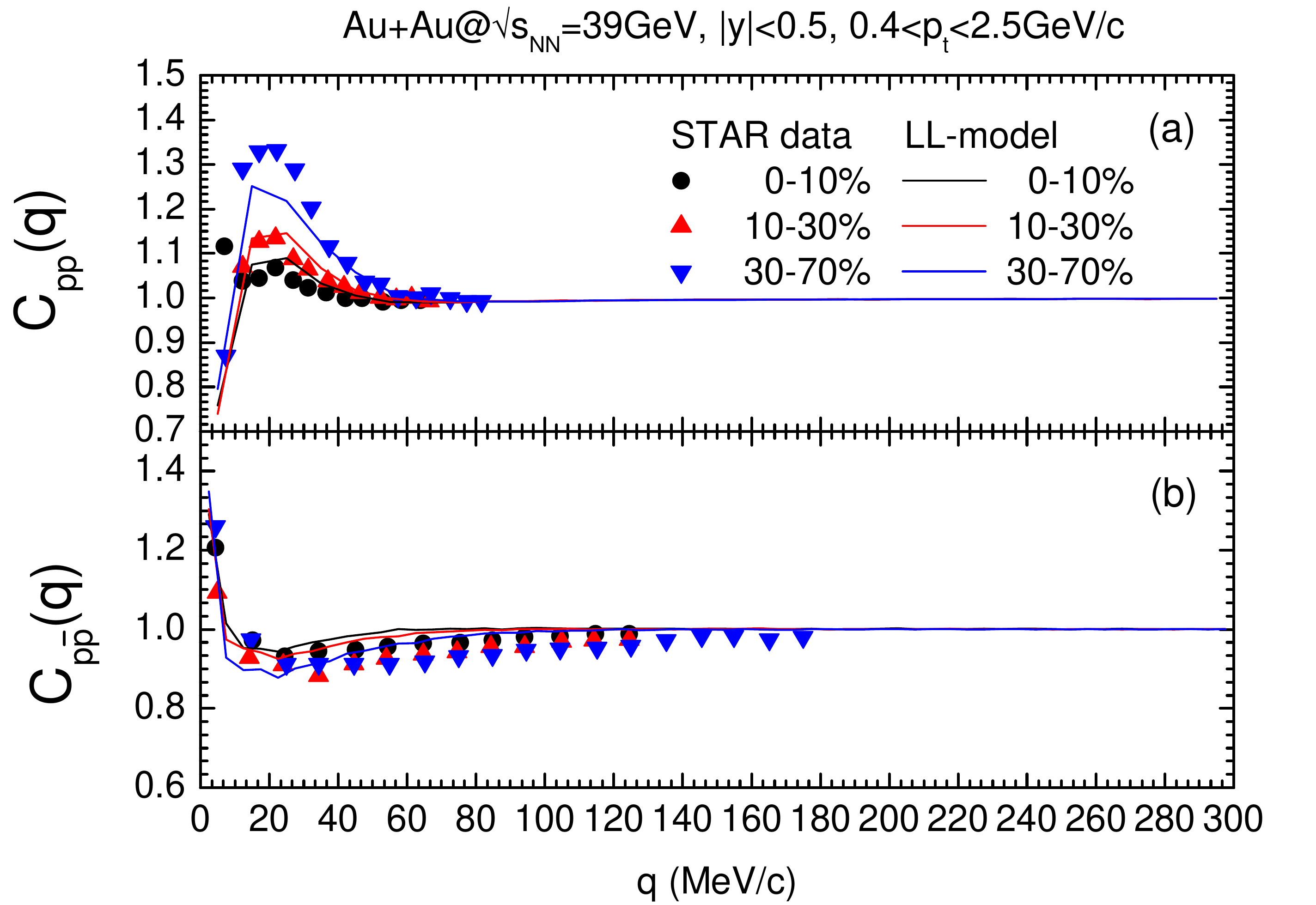}
% \includegraphics[width=\linewidth]{fig1-pp-ppbar-exp.png}
 %\vspace{-18cm}
 \centering
 \caption{
Proton-proton (a) and proton-antiproton (b) momentum correlation functions for different centrality classes in $\sqrt{s_{NN}}$ = 39 GeV Au + Au collisions. Solid markers represent the preliminary experimental data from the RHIC-STAR collaboration~\cite{Zbroszczyk2019,Siejka2019}, and lines represent our model calculation results from the AMPT model plus the Lednick$\acute{y}$ and Lyuboshitz code.  Note that the longer hadronic rescattering time of 700 $fm/c$ is used in this specific calculation for comparing with the data.
}
 \label{fig1}
\end{figure}

Fig.~\ref{fig1} presents results of $p$-$p$ and $p$-$\bar{p}$ correlation functions for three different centrality classes of $0-10$ $\%$, $10-30$ $\%$, and $30-70$ $\%$ calculated by the AMPT model in Au + Au collisions at $\sqrt{s_{NN}}$ = 39 GeV. Within the cut of  transverse momentum $p_{t}$ and  rapidity $y$, we confront the experimental data with the predictions of the AMPT model combined with Lednick$\acute{y}$ and Lyuboshitz code. When the phase-space information of nucleons at the maximum  rescattering time  among hadrons of 700 $fm/c$ is selected from the AMPT model, it is found that the results can well describe the experimental data for the momentum correlation functions of $p$-$p$ and $p$-$\bar{p}$ from the RHIC-STAR collaboration~\cite{Zbroszczyk2019,Siejka2019}, especially in more central collisions. Considering that the preliminary experimental results were not corrected by feed-down effect corrections~\cite{Zbroszczyk2019,Siejka2019}, the real correlation functions for primary $p$-$p$ and  $p$-$\bar{p}$  could be much more stronger. In this case, using much longer MRT of 700 $fm/c$ in the AMPT model might be a reasonable choice for making quantitative comparison with feed-down uncorrected data since the system will become more expanded and weakly correlated among particles after longer MRT in AMPT. However, the quantitative reproduction is not our main concern in the present work. In the following calculations, we fixed the MRT at 100 fm/c and presented systematic results among different light (anti)nuclei.  However, as  one can notice
that the results for $p$-$p$ and $p$-$\bar{p}$  change substantially when changing the MRT by comparing Fig.~\ref{fig1} and ~\ref{fig2}. To estimate this uncertainty, we also check some results for light nuclei correlations with different MRT. For example,  $d$-$d$ or $p$-$d$ correlations for MRT equal 
700fm/c. It is found that the correlation becomes  slightly weaker  at smaller $q$ (i.e. a little larger value of $C_{dd}$ or $C_{pd}$ close to 1 at MRT = 700 fm/c), which has the similar trend as $p$-$p$ and $p$-$\bar{p}$ cases.  
But the uncertainty is less than  20\%  at the lowest relative momentum  and tends to vanish at $q >$ 50 MeV/c for light nuclei correlations ($d$-$d$ or $p$-$d$)  when changing the MRT from 100fm/c to 700fm/c, which can be essentially understood by weak feed-down effects for light nuclei.  In addition, we also check the $p$-$d$ correlation with different  velocity selection. Only less than 10\% uncertainty is found for lower $q$ between the case of MRT equal 700 fm/c to the one at 100fm/c.
By this comparison of results at MRT equal 700 and 100 fm/c, we conclude that nucleon-(anti)nucleon correlations are much  influenced by the MRT but light nuclei correlations only change slightly. Overall, the MRT = 100 fm/c is basically safe choice for such light nuclei correlations.

 %%%%%%%%%%%%%%%%%%%%%%%%%%%%%%%%%%%%%%%fig1%%%%%%%%%%%%%%%%%%%%%%%%%%
%%%%%%%%%%%%%%%%%%%%%%%%%%%%%%%%%%%%%%%%%%%%%%%%%%%%%%%%%%%%%%%%%%%%
\subsection{Centrality and system-size dependence of identical light (anti)nuclei momentum correlation functions }

\begin{figure}[!htbp]
 \includegraphics[width=\linewidth]{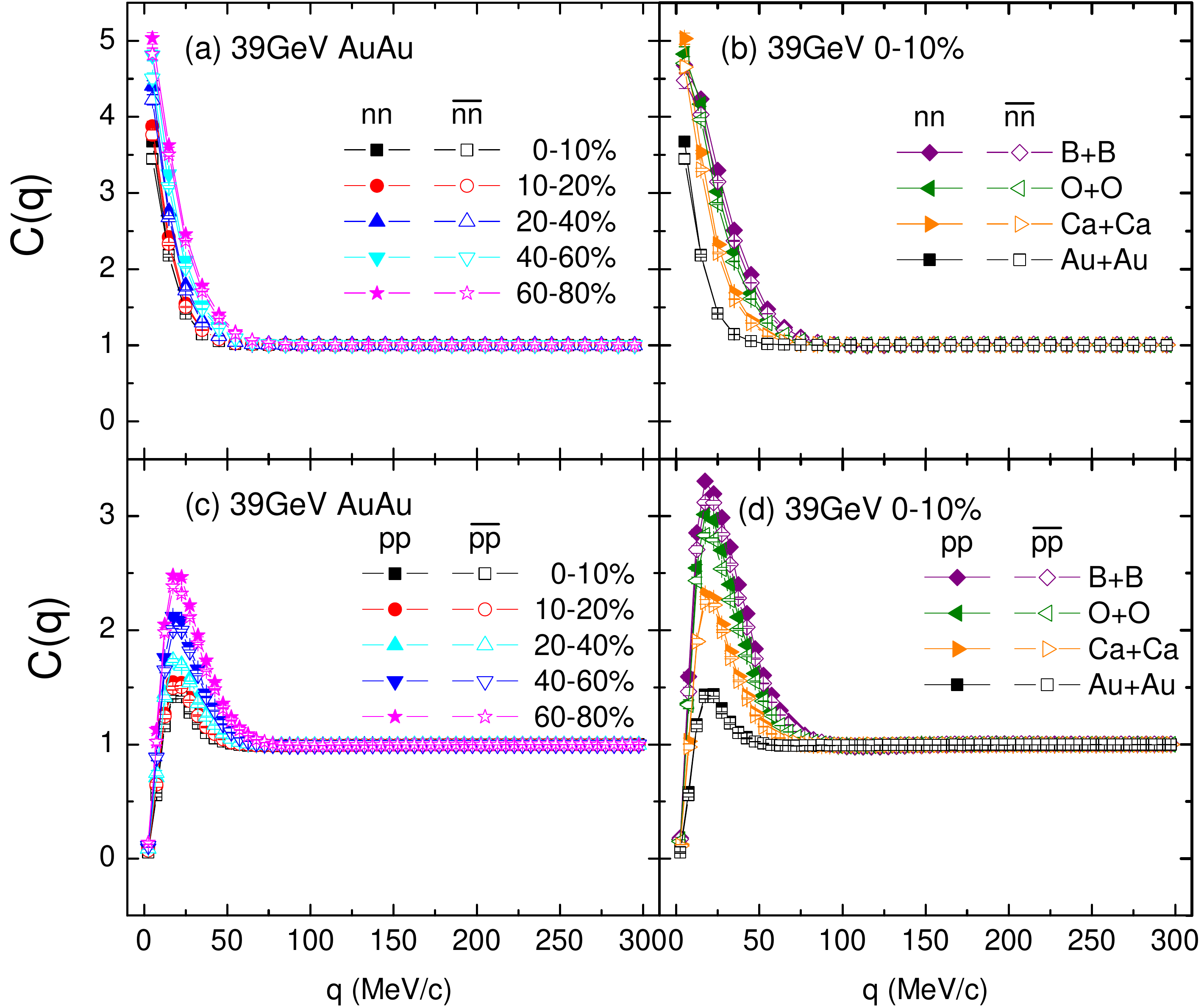}
%  \includegraphics[width=\linewidth]{fig2-nn-pp-B-Au.png}
%\vspace{-10cm}
 \centering
 \caption{
The momentum correlation functions at mid-rapidity  ($\left|y \right|< 0.5$)  of (anti)neutron-pairs and (anti)proton-pairs as a function of five different centralities for $_{79}^{197}\textrm{Au}+_{79}^{197}\textrm{Au}$ reaction at $\sqrt{s_{NN}}$ = 39 GeV are presented in (a) and (c), respectively. The momentum correlation functions of (anti)neutron-pairs and (anti)proton-pairs at mid-rapidity ($\left|y \right|< 0.5$) for 0-10$\%$ central collisions of $_{5}^{10}\textrm{B}+_{5}^{10}\textrm{B}$, $_{8}^{16}\textrm{O}+_{8}^{16}\textrm{O}$, $_{20}^{40}\textrm{Ca}+_{20}^{40}\textrm{Ca}$ as well as $_{79}^{197}\textrm{Au}+_{79}^{197}\textrm{Au}$ systems at $\sqrt{s_{NN}}$ = 39 GeV are presented in (b) and (d), respectively. The $p$-$p$ and $n$-$n$ momentum correlation functions (solid symbols) and the anti-one (open symbols) are shown in each panel.
 }
 \label{fig2}
\end{figure}

The centrality dependence of the two-particle momentum correlation function can systematically investigate the contributions from the system-size and particle interactions on the correlations.
Fig.~\ref{fig2} (a) and (c) present the momentum correlation functions of identical (anti)particle pairs ($n$-$n$ ($\bar{n}$-$\bar{n}$) and $p$-$p$ ($\bar{p}$-$\bar{p}$) ) for $_{79}^{197}\textrm{Au}+_{79}^{197}\textrm{Au}$ collisions at different centralities of $0-10$ $\%$, $10-20$ $\%$, $20-40$ $\%$, $40-60$ $\%$, and $60-80$ $\%$ at $\sqrt{s_{NN}}$ = 39 GeV. The momentum correlation functions of (anti)neutron pairs exhibit more than unity in Fig.~\ref{fig2} (a), which is caused by the attractive $s$-wave interaction between the two (anti)neutrons. In Fig.~\ref{fig2} (c), the shape of the (anti)proton$-$(anti)proton momentum correlation functions looks as expected from the interplay between the quantum statistical (QS) and final state interactions (FSI) and is consistent with previous 
results~\cite{Star-nature,wtt2018,wtt2019}. The (anti)proton$-$(anti)proton  momentum correlation functions exhibit less than unity at low relative momentum $q$ in Fig.~\ref{fig2} (c), which is mainly caused by the Coulomb repulsion between the (anti)proton pairs. With increasing relative momentum, the attractive $s$-wave interaction between the two (anti)protons gives rise to a maximum of the (anti)proton$-$(anti)proton momentum correlation functions at $q$ around 0.020 GeV in Fig.~\ref{fig2} (c). The antiproton$-$antiproton momentum correlation functions show a similar structure with proton pairs, resulting from the same attractive interaction between two antiprotons 
~\cite{Star-nature}. Fig.~\ref{fig2} (a) and (c) compare five centralities of $0-10$ $\%$, $10-20$ $\%$, $20-40$ $\%$, $40-60$ $\%$, and $60-80$ $\%$ of the two-(anti)particle momentum correlation functions. The enhanced strength of the $n$-$n$ ($\bar{n}$-$\bar{n}$) and $p$-$p$ ($\bar{p}$-$\bar{p}$) momentum correlation functions is observed in peripheral collisions. These results indicate that (anti)particle emission occurs from a source with smaller space  extent in peripheral collision. In addition, the effect of system$-$size on the momentum correlation functions of (anti)particles is also investigated by four different systems, namely $_{5}^{10}\textrm{B}+_{5}^{10}\textrm{B}$, $_{8}^{16}\textrm{O}+_{8}^{16}\textrm{O}$, $_{20}^{40}\textrm{Ca}+_{20}^{40}\textrm{Ca}$ and $_{79}^{197}\textrm{Au}+_{79}^{197}\textrm{Au}$, in central collisions. In Fig.~\ref{fig2} (b) and (d), the $n$-$n$ ($\bar{n}$-$\bar{n}$) and $p$-$p$ ($\bar{p}$-$\bar{p}$) momentum correlation functions appear strong sensitivity to system-size and an enhanced strength is observed when particle pairs are emitted from smaller system collisions. This  enhanced strength of the momentum correlation functions for particle pairs is a physical effect stemming from the smaller space extent of the emission source~\cite{Ghetti2000}. Therefore, the emission source-size of particle pairs obtained by their momentum correlation functions and system-size is self-consistent. 

%%%%%%%%%%%%%%%%%%%%%%%%%%%%%%%%%%%%%%%fig2%%%%%%%%%%%%%%%%%%%%%%%%%%
%%%%%%%%%%%%%%%%%%%%%%%%%%%%%%%%%%%%%%%%%%%%%%%%%%%%%%%%%%%%%%%%%%%%
%%%%%%%%%%%%%%%%%%%%%%%%%%%%%%%%%%%%%%%fig3%%%%%%%%%
\begin{figure}[!htbp]
 \includegraphics[width=\linewidth]{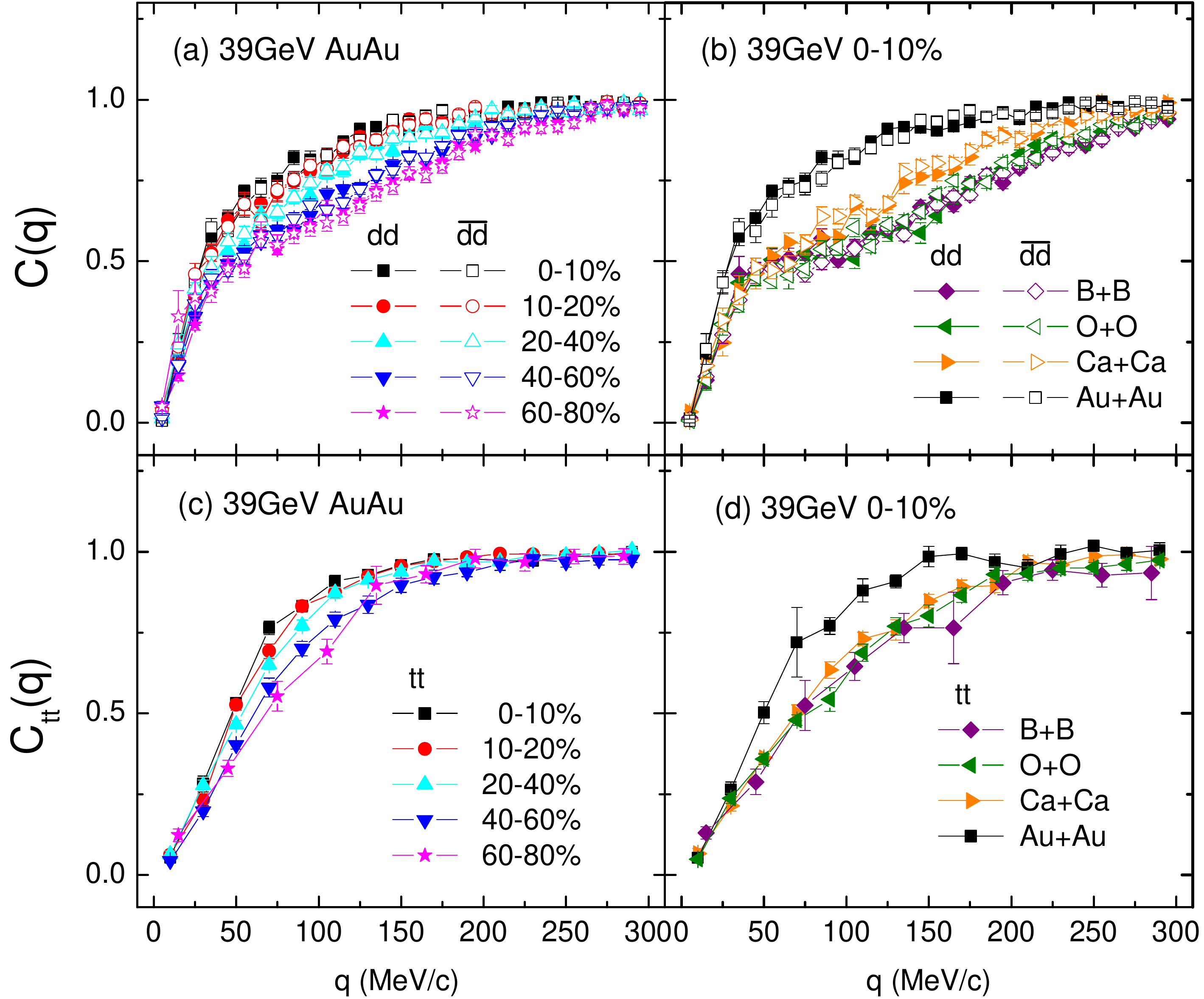}
 \centering
 \caption{
Same as Fig.~\ref{fig2} but for the light (anti)cluster pairs. (a) and (b) for $d-d$ momentum correlation functions (solid symbols) and the anti-one (open symbols), (c) and (d) for  $t-t$ momentum correlation functions (solid symbols). 
 }
 \label{fig3}
\end{figure}
%%%%%%%%%%%%%%%%%%%%%%%%%%%%%%%%%%%%%%%%%%%%%%%%%%

%%%%%%%%%%%%%%%%%%%%%%%%%%%%%%%%%%%%%%%fig4%%%%%%%%%%%%%%%%%%%%%%%%%%
\begin{figure}[!htbp]
 \includegraphics[width=\linewidth]{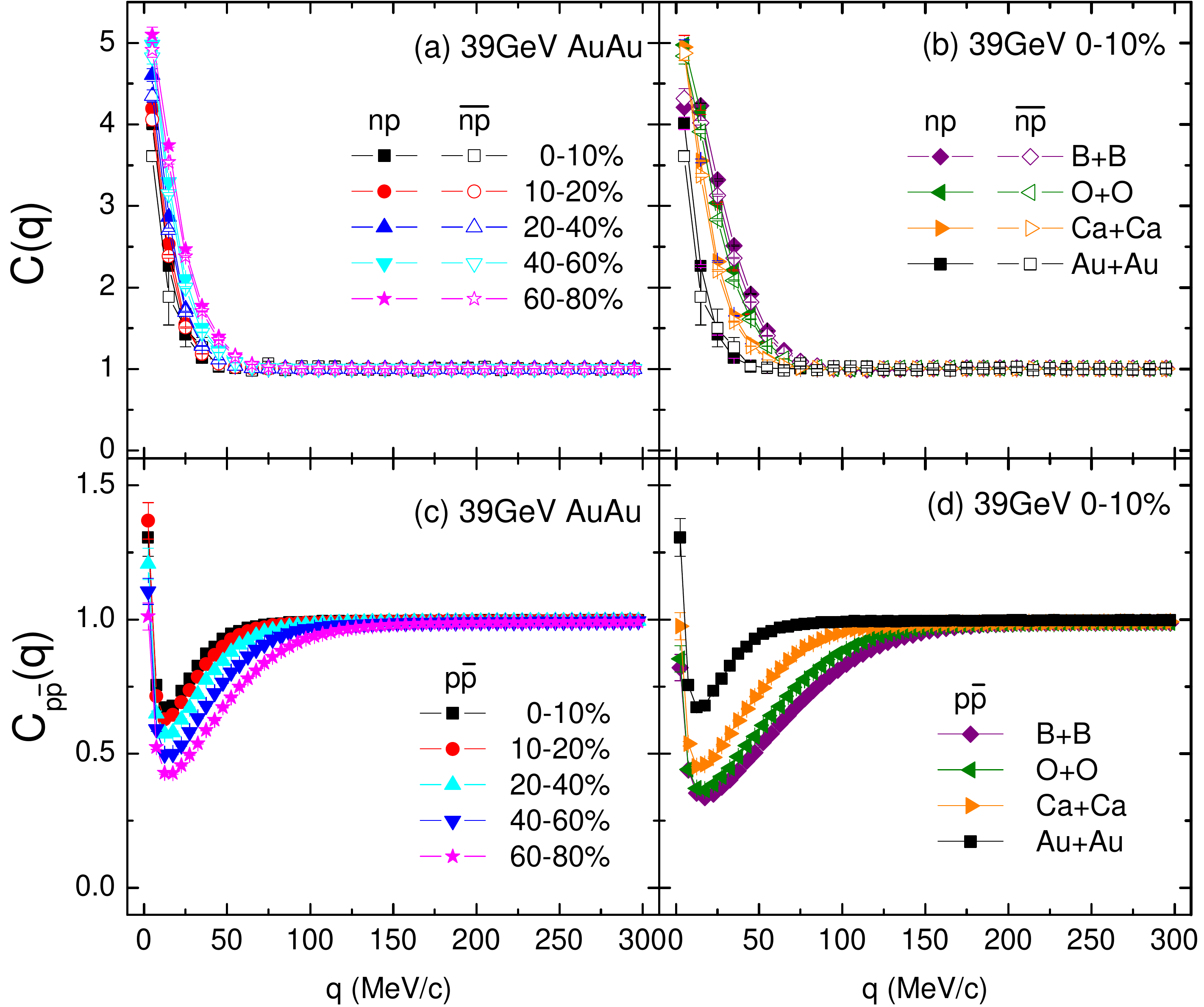}
 \centering
 \caption{
Same as Fig.~\ref{fig2} but for the nonidentical particle pairs. (a) and (b) for $n$-$p$ momentum correlation functions (solid symbols) and the anti-one (open symbols), (c) and  (d for) $p$-$\bar{p}$ momentum correlation functions (solid symbols).
 }
 \label{fig4}
\end{figure}
%%%%%%%%%%%%%%%%%%%%%%%%%%%%%%%%%%%%%%%%%%%%%%%%%%%%%%%%%%%%%%%%%%%%

%%%%%%%%%%%%%%%%%%%%%%%%%%%%%%%%%%%%%%%fig5%%%%%%%%%
\begin{figure*}[!htbp]
 \includegraphics[width=0.8\linewidth]{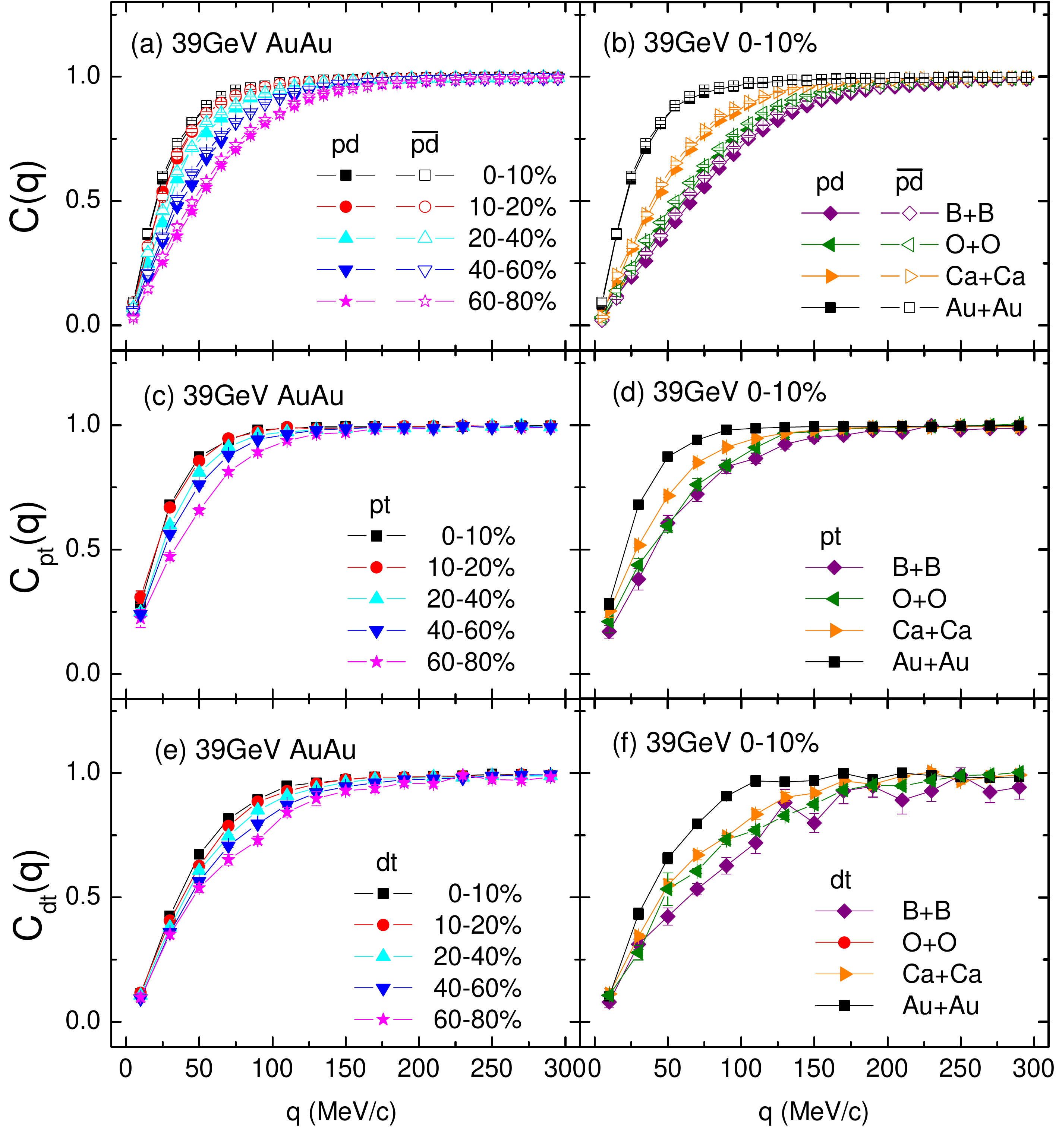}
 \centering
 \caption{
Same as Fig.~\ref{fig4} but for nonidentical light (anti)nuclei. (a) and (b) for $p$-$d$ momentum correlation functions (solid symbols) and the anti-one (open symbols). (c) and (d) for  $p$-$t$ momentum correlation functions (solid symbols), (e) and (f) for $d$-$t$ momentum correlation functions (solid symbols). 
 }
 \label{fig5}
\end{figure*}

%%%%%%%%%%%%%%%%%%%%%%%%%%%%%%%%%%%%%%%fig6%%%%%%%%%%%%%%%%%%%%%%%%%%
\begin{figure}[!htbp]
 \includegraphics[width=\linewidth]{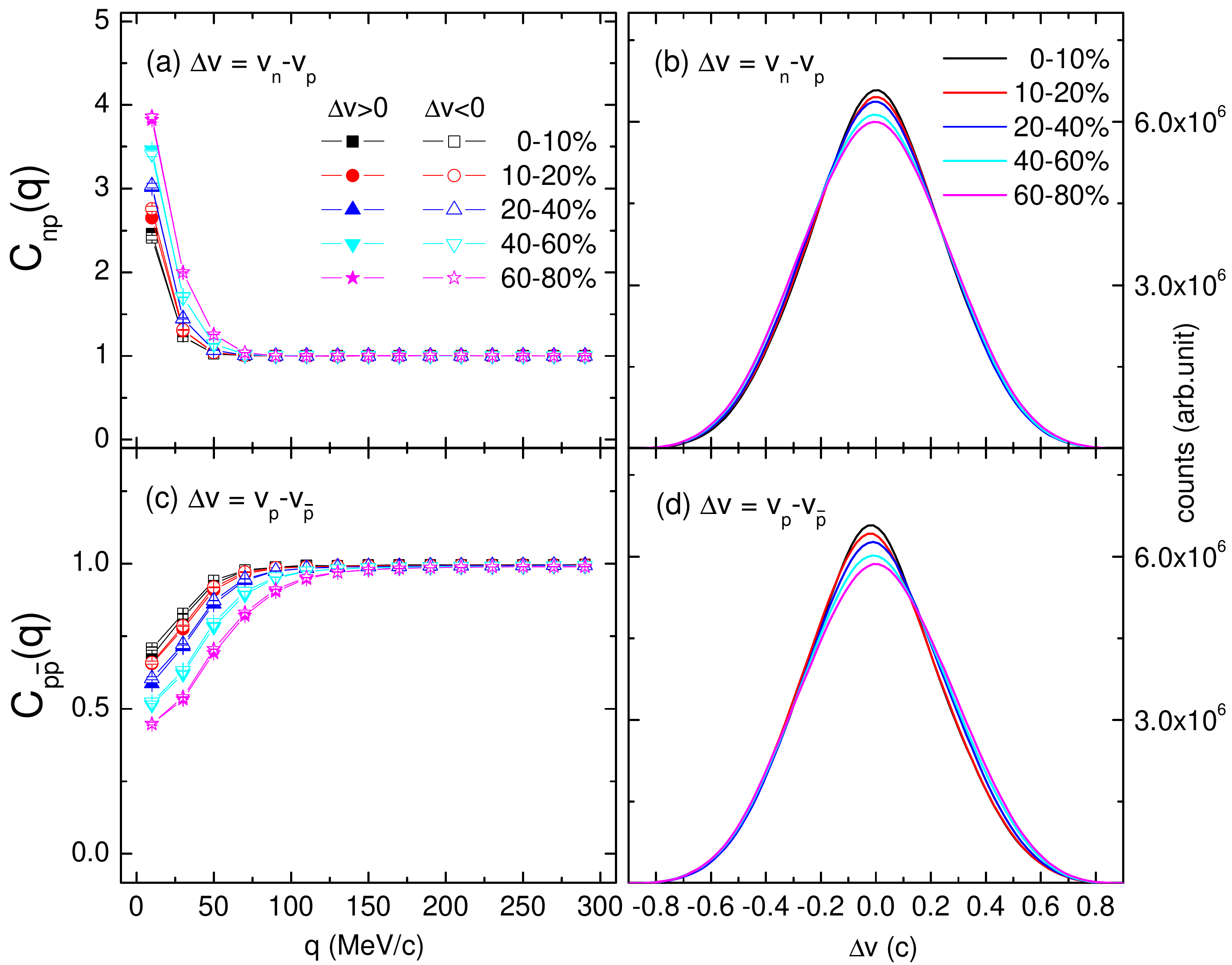}
 \centering
 \caption{
The velocity-gated momentum correlation functions (left) and velocity difference ($\Delta v$) spectra (right) for $n$-$p$ and $p$-$\bar{p}$ as a function of five different centralities in mid-rapidity ($\left|y \right|< 0.5$) for  39 GeV $_{79}^{197}\textrm{Au}+_{79}^{197}\textrm{Au}$ collision. The velocity conditions are indicated in each panel: $\Delta v >0$ is remarked by solid symbol and the $\Delta v < 0$ by open symbol.
}
 \label{fig6}
\end{figure}
%%%%%%%%%%%%%%%%%%%%%%%%%%%%%%%%%%%%%%%%%%%%%%%%%%%%%%%%%%%%%%%%%%%%

%%%%%%%%%%%%%%%%%%%%%%%%%%%%%%%%%%%%%%%fig7%%%%%%%%%%%%%%%%%%%%%%%%%%
\begin{figure}[!htbp]
 \includegraphics[width=\linewidth]{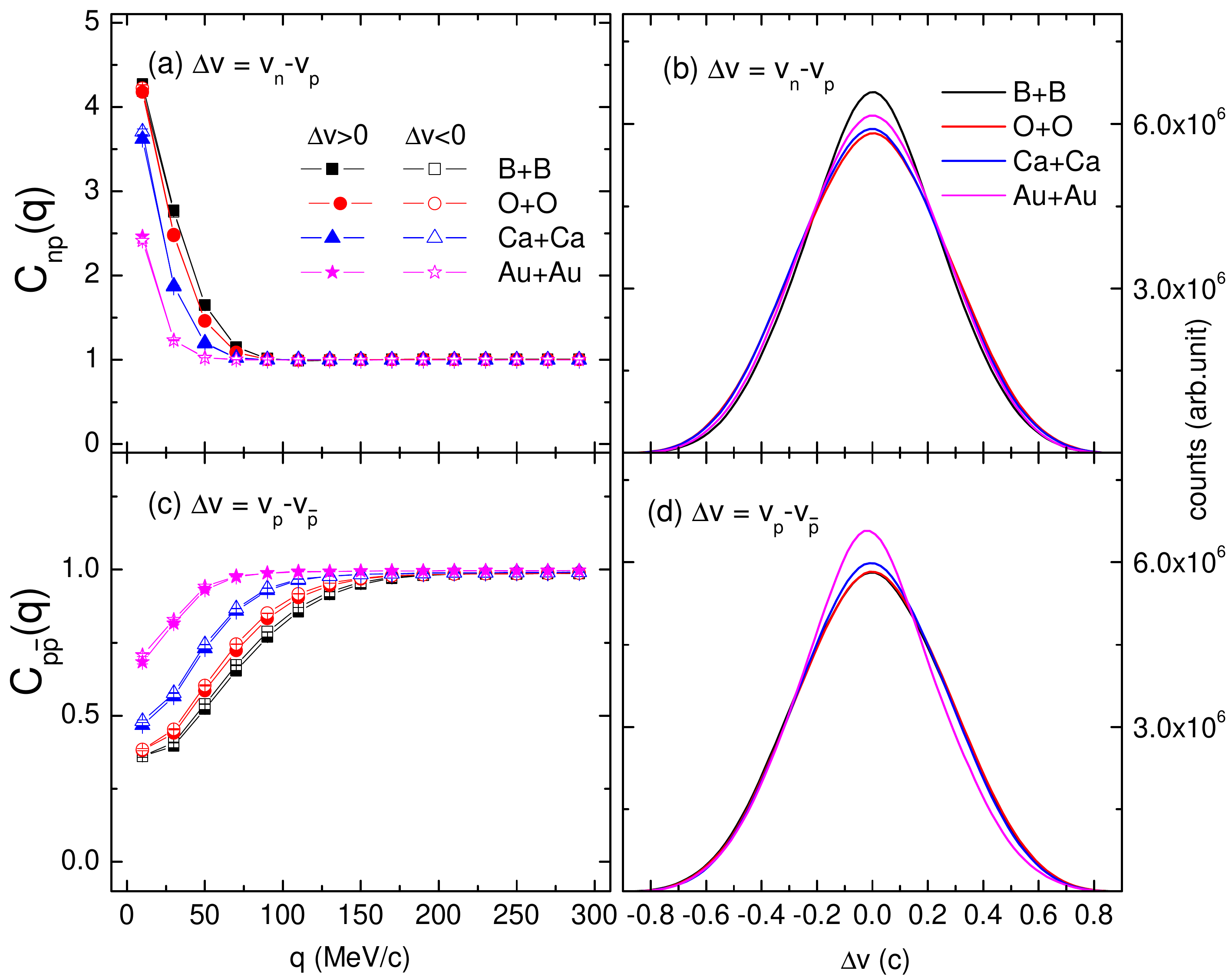}
 \centering
 \caption{
Same as Fig.~\ref{fig6} but for 0 $-$ 10 $\%$ central collisions of $_{5}^{10}\textrm{B}+_{5}^{10}\textrm{B}$, $_{8}^{16}\textrm{O}+_{8}^{16}\textrm{O}$, $_{20}^{40}\textrm{Ca}+_{20}^{40}\textrm{Ca}$ as well as $_{79}^{197}\textrm{Au}+_{79}^{197}\textrm{Au}$ systems at $\sqrt{s_{NN}}$ = 39 GeV. The velocity conditions are indicated in each panel: $\Delta v >0$ is remarked by solid symbol and the $\Delta v < 0$ by open symbol.}
 \label{fig7}
\end{figure}
%%%%%%%%%%%%%%%%%%%%%%%%%%%%%%%%%%%%%%%%%%%%%%%%%%%%%%%%%%%%%%%%%%%%
%%%%%%%%%%%%%%%%%%%%%%%%%%%%%%%%%%%%%%%fig8%%%%%%%%%%%%%%%%%%%%%%%%%%
\begin{figure}[!htbp]
 \includegraphics[width=\linewidth]{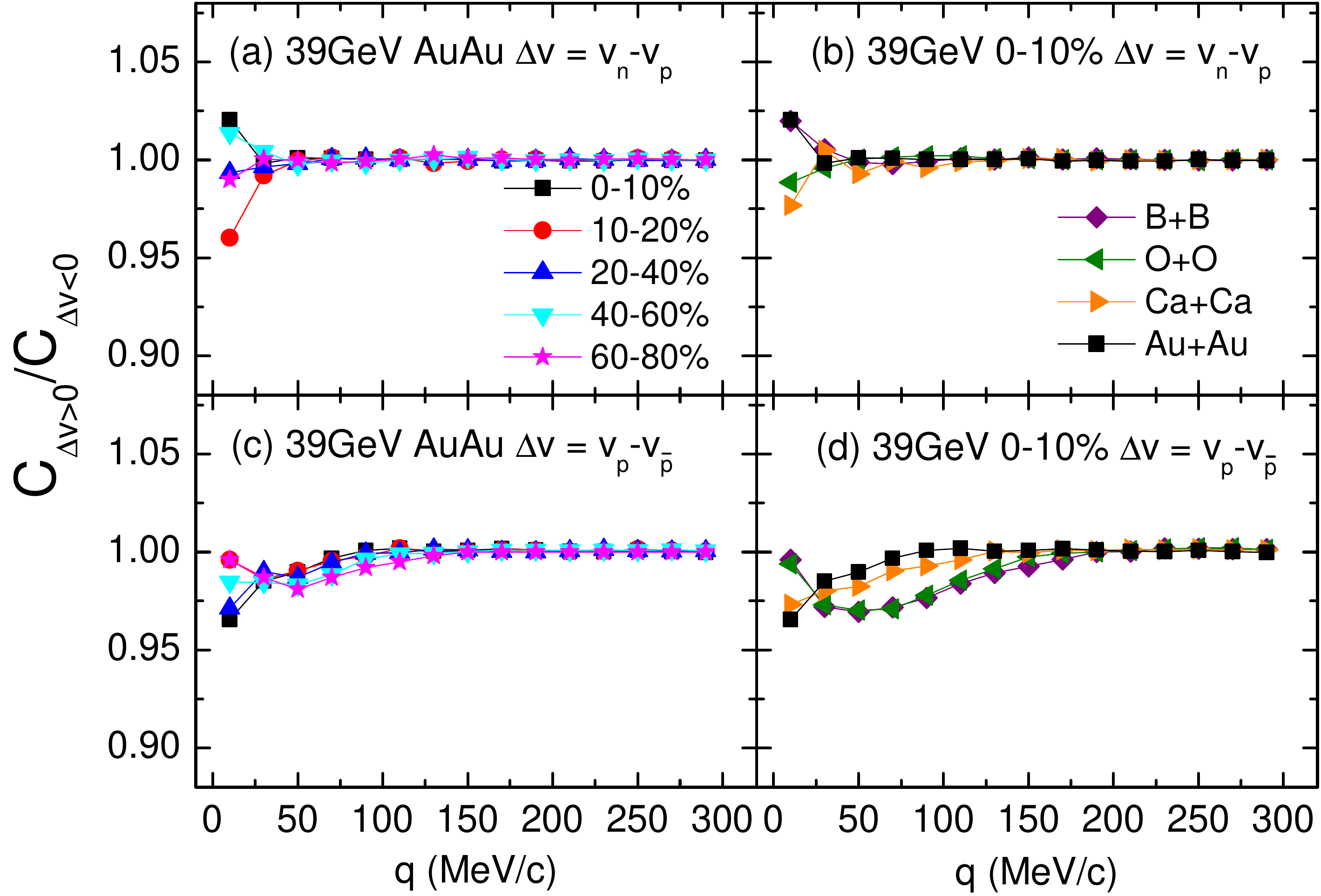}
 \centering
 \caption{
Ratios of the velocity-gated momentum correlation functions (left) of $n$-$p$ (a) and $p$-$\bar{p}$ (c) pairs  for 39 GeV $_{79}^{197}\textrm{Au}+_{79}^{197}\textrm{Au}$ collision at  mid-rapidity ($\left|y \right|< 0.5$)  and five different centralities. Ratios of the velocity-gated momentum correlation functions (right) of $n$-$p$ (b) and $p$-$\bar{p}$ (d) pairs for 0 $-$ 10 $\%$ central collisions of $_{5}^{10}\textrm{B}+_{5}^{10}\textrm{B}$, $_{8}^{16}\textrm{O}+_{8}^{16}\textrm{O}$, $_{20}^{40}\textrm{Ca}+_{20}^{40}\textrm{Ca}$ as well as $_{79}^{197}\textrm{Au}+_{79}^{197}\textrm{Au}$ systems at $\sqrt{s_{NN}}$ = 39 GeV.
 }
 \label{fig8}
\end{figure}
%%%%%%%%%%%%%%%%%%%%%%%%%%%%%%%%%%%%%%%%%%%%%%%%%%%%%%%%%%%%%%%%%%%%

%%%%%%%%%%%%%%%%%%%%%%%%fig9%%%%%%%%%%%%%%%%%%%%%%%%%%
\begin{figure*}[!htbp]
 \includegraphics[width=0.8\linewidth]{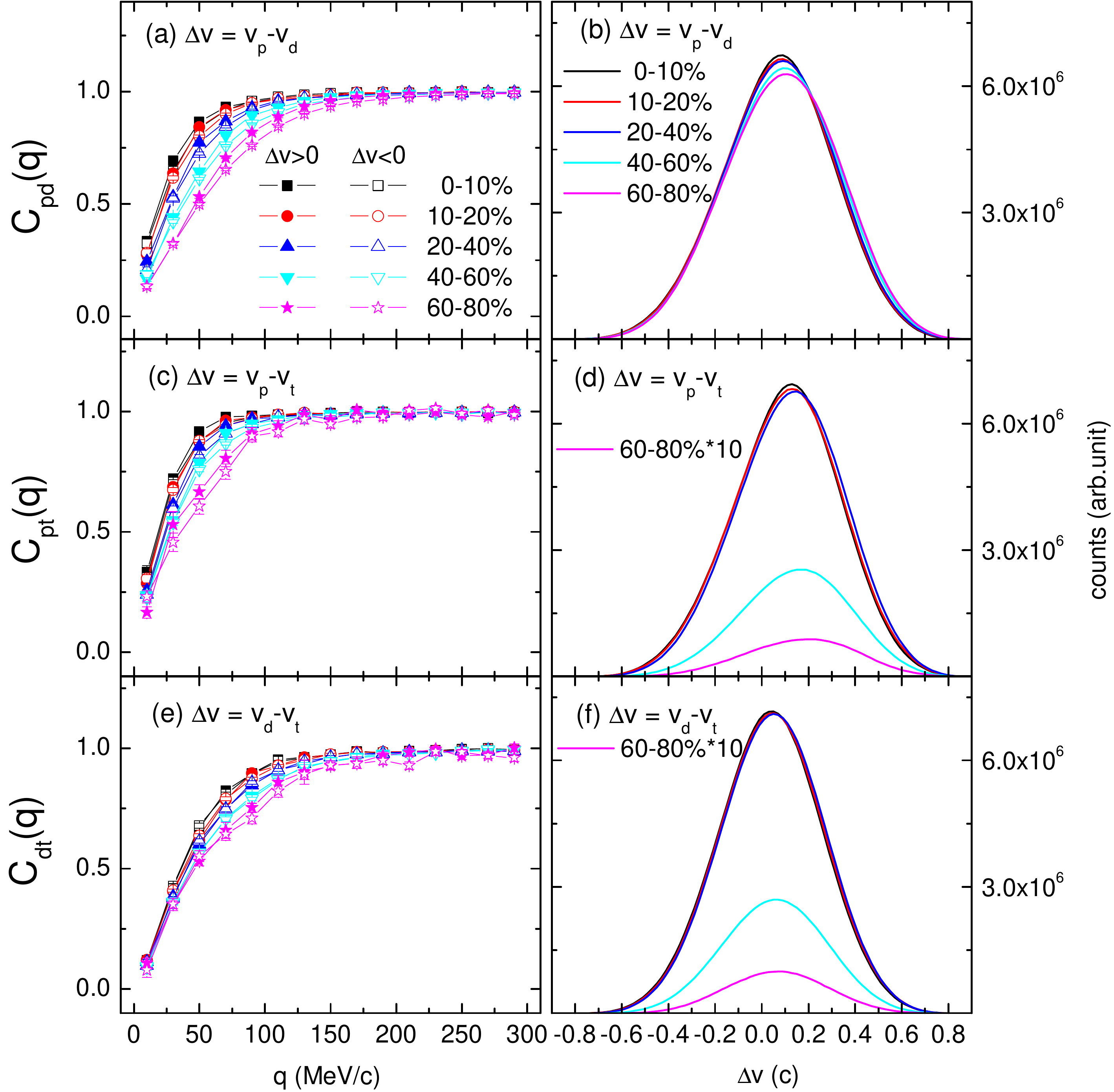}
 \centering
 \caption{
 Same as Fig.~\ref{fig6} but for $p$-$d$ (a) and (b), $p$-$t$ (c) and (d), and $d$-$t$ (e) and (f) pairs. 
 }
 \label{fig9}
\end{figure*}

%%%%%%%%%%%%%%%%%%%%%%%%%%%%%%%%%%%%%%%%%%%%%%%%%%%%%%%%%%%%%%%%%%%%

%%%%%%%%%%%%%%%%%%%%%%%%fig10%%%%%%%%%%%%%%%%%%%%%%%%%%
\begin{figure*}[!htbp]
 \includegraphics[width=0.8\linewidth]{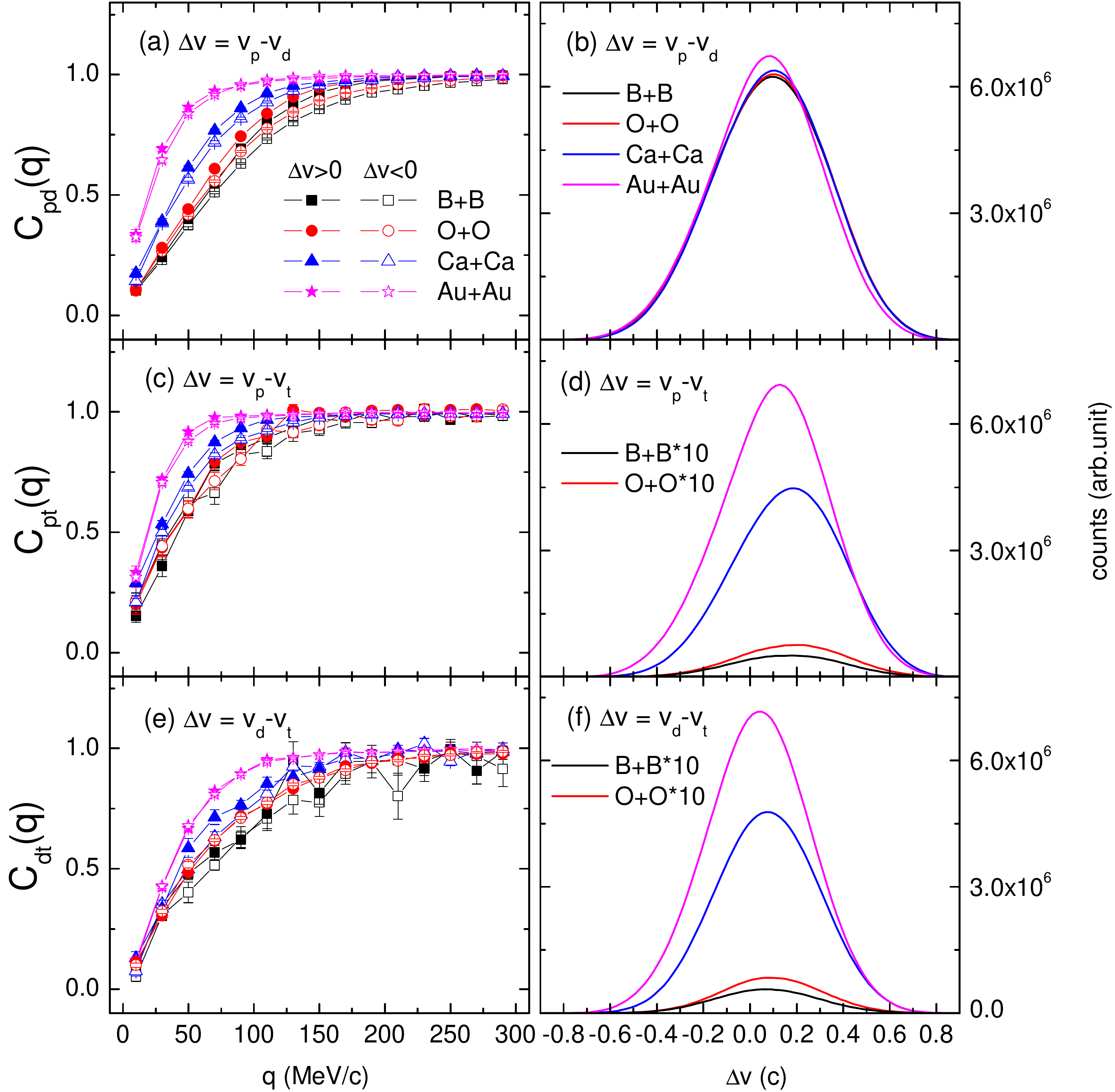}
 \centering
 \caption{
Same as Fig.~\ref{fig7} but for $p$-$d$ (a) (b), $p$-$t$ (c) (d) and $d$-$t$ (e) (f) pairs. 
 }
 \label{fig10}
\end{figure*}

%%%%%%%%%%%%%%%%%%%%%%%%%%%%%%%%%%%%%%%%%%%%%%%%%%%%%%%%%%%%%%%%%%%%

%%%%%%%%%%%%%%%%%%%%%%%%fig11%%%%%%%%%%%%%%%%%%%%%%%%%%
\begin{figure*}[!htbp]
 \includegraphics[width=0.8\linewidth]{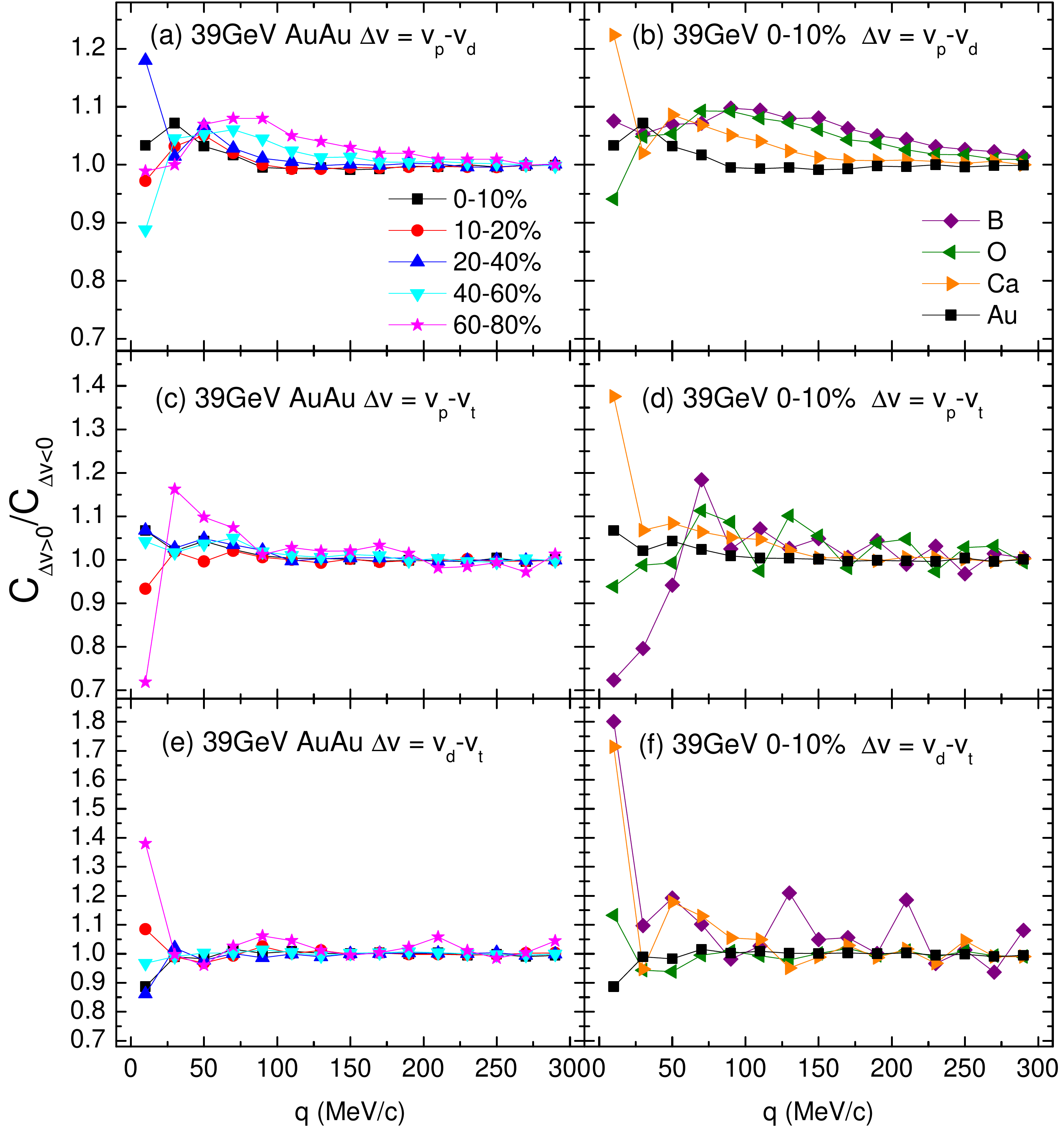}
 \centering
 \caption{
Same as Fig.~\ref{fig8} but for $p$-$d$ (a) and (b), $p$-$t$ (c) and (d) and $d$-$t$ (e) and (f) pairs. 
 }
 \label{fig11}
\end{figure*}

%%%%%%%%%%%%%%%%%%%%%%%%%%%%%%%%%%%%%%%%%%%%%%%%%%%%%%%%%%%%%%%%%%%%

%%%%%%%%%%%%%%%%%%%%%%%%%%%%%%%%%%%%%%%%%%%%%%%%%%%%%%%%%%%%%%%%%%%%

Figure~\ref{fig3} shows the centrality and system-size dependences of the momentum correlation functions for light (anti)cluster in similar condition as in Fig.~\ref{fig2}. Figure~\ref{fig3} (a) and (c) present the momentum correlation functions of $d$-$d$ ($\bar{d}$-$\bar{d}$) and $t$-$t$ for $_{79}^{197}\textrm{Au}+_{79}^{197}\textrm{Au}$ collisions at different centralities of $0-10$ $\%$, $10-20$ $\%$, $20-40$ $\%$, $40-60$ $\%$, and $60-80$ $\%$ at $\sqrt{s_{NN}}$ = 39 GeV. The $d$-$d$ ($\bar{d}$-$\bar{d}$) momentum correlation functions exhibit less than unity at lower relative momentum $q$ in Fig.~\ref{fig3} (a) and (b), which is caused by the Coulomb repulsion. The two-triton momentum correlation functions are less than unity with increasing relative momentum $q$ as shown in Fig.~\ref{fig2} (c) and (d), which is caused by only the Coulomb potential in the Lednick$\acute{y}$ and Lyuboshitz code~\cite{lednicky2006,lednicky2009,lednicky2008}. The antideuteron$-$antideuteron momentum correlation function also shows an exact similar shape with deuteron pairs due to the similar phase-space distributions between deuteron and antideuteron.
Due to significant less yields of tritons which induce too large error, the antitriton$-$antitriton momentum correlation function is not shown in the present work, which should be observed as the same trend with triton pairs. Fig.~\ref{fig3} (a) and (c) also compare five centralities of $0-10$ $\%$, $10-20$ $\%$, $20-40$ $\%$, $40-60$ $\%$, and $60-80$ $\%$ for the momentum correlation functions of two light (anti)clusters. The larger suppression of the $d$-$d$ ($\bar{d}$-$\bar{d}$) and $t$-$t$ correlation functions is clearly visible in peripheral collisions. These results also indicate that light (anti)cluster emission occurs from a source with smaller space extent for peripheral collision, which is similar to  Fig.~\ref{fig2} (a) and (c). In Fig.~\ref{fig3} (b) and (d), an enhanced strength of the momentum correlation function for $d$-$d$ ($\bar{d}$-$\bar{d}$) and $t$-$t$ is also observed when light (anti)cluster pairs emitted from smaller systems, such as in B + B and O + O collisions. However,  the sensitivity seems disappear  in these small systems.

\subsection{Nonidentical light (anti)nuclei momentum correlation functions gated on centrality and system-size}

Now we  investigate centrality and system-size dependence of the nonidentical (anti)particle momentum correlation functions, such as $n$-$p$ ($\bar{n}$-$\bar{p}$), $p$-$\bar{p}$, $p$-$d$ ($\bar{p}$-$\bar{d}$), $p$-$t$ and $d$-$t$. Fig.~\ref{fig4} (a) and (c) show results for the momentum correlation functions of $n$-$p$ ($\bar{n}$-$\bar{p}$) and $p$-$\bar{p}$ for the same centrality classes as Fig.~\ref{fig2}. The same centrality dependence is also clearly seen in Fig.~\ref{fig4} (a) and (c). Because of the strong attractive final state interaction between $n$ and $p$, the $n$-$p$ ($\bar{n}$-$\bar{p}$) momentum correlation functions show a strong positive correlation at small values of the relative momentum $q$ in Fig.~\ref{fig4} (a) and (b). Fig.~\ref{fig4} (c) shows results for proton$-$antiproton momentum correlation functions, which are different from the results for proton pairs in Fig.~\ref{fig2} (c), however, qualitatively agrees with the experimental results in Ref.~\cite{Zbroszczyk2019,Siejka2019}.
In addition, Fig.~\ref{fig4} (b) and (d) show system-size dependence of $n$-$p$ ($\bar{n}$-$\bar{p}$) and $p$-$\bar{p}$ momentum correlation functions, which is almost unanimously with the identical (anti)particle one in Fig.~\ref{fig2} (b) and (d). We can also observe an enhanced strength of  momentum correlation function for particle pairs in smaller systems. In the same way, we also investigate the effect of different centralities and system-size on the momentum correlation functions of nonidentical light (anti)nuclei. The $p$-$d$ ($\bar{p}$-$\bar{d}$), $p$-$t$ and $d$-$t$ momentum correlation functions in Fig.~\ref{fig5} (a), (c) and (e) are all characterized by an anti-correlation feature. For the $p$-$d$ ($\bar{p}$-$\bar{d}$) momentum correlation functions in Fig.~\ref{fig5} (a), the anti-correlation shape is a little unlike to the proton$-$deuteron momentum correlation function in the intermediate energy heavy-ion collision~\cite{wtt2019,YijieWang2022}, indicating that a competition between the $s$-wave attraction and the Coulomb repulsion. The correlation functions of $p$-$t$ and $d$-$t$ in Fig.~\ref{fig4} (c) and (e) also display the trend of below unity due to the dominant Coulomb repulsion, which is similar to the previous results in intermediate energy heavy-ion collisions~\cite{wtt2019,YijieWang2022}. In Fig.~\ref{fig5} (b), the system-size dependence of $p$-$d$ ($\bar{p}$-$\bar{d}$) momentum correlation functions is shown,  an enhancement of  $p$-$d$ ($\bar{p}$-$\bar{d}$) momentum correlation function is observed in smaller systems. In Fig.~\ref{fig5} (d) and (f), the $p$-$t$ and $d$-$t$ momentum correlation functions appear more sensitive to system-size only in the large system such as Au and Ca. 

\subsection{Velocity selected nonidentical light nuclei momentum correlation functions}

The momentum correlation functions of unlike particles can provide an independent constrain on their mean emission order by simply making velocity selections~\cite{Gelderloos1994,Gelderloos1995,DGourio2000,Lednicky1996,HuangBS1}.  The principle of comparing the velocity-gated momentum correlation functions for the nonidentical particle pair to infer their average emission order is as follows. Here the two nonidentical particles are named by “a” and “b”, respectively. If the velocity of “a” particle is lower than “b” particle, the (anti)correlation will be stronger when the “a” particle is emitted averagely early than the “b” particle, because the space-size between them is reduced during the flight and the final-state interaction (FSI) is enhanced, and vice versa. In addition, the velocity difference ($\Delta$v) spectrum between the two nonidentical particles is also sensitive to the mean emission order.
Fig.~\ref{fig6} presents the velocity-gated momentum correlation functions as well as velocity difference ($\Delta$v) spectra of unlike particles pairs $n$-$p$ and $p$-$\bar{p}$ for 39 GeV $_{79}^{197}\textrm{Au}+_{79}^{197}\textrm{Au}$ collisions at different centralities of $0-10$ $\%$, $10-20$ $\%$, $20-40$ $\%$, $40-60$ $\%$, and $60-80$ $\%$. In Fig.~\ref{fig6} (a) and (c), the centrality dependence on the velocity-gated momentum correlation functions of $n$-$p$ and $p$-$\bar{p}$ is similar to Fig.~\ref{fig4}. In Fig.~\ref{fig6} (a), the momentum correlation function for $n$-$p$ pair with $v_{n}$ $>$ $v_{p}$ is similar to one with the reverse situation. The symmetry of velocity difference ($\Delta$v) spectra for $n$-$p$ pairs is shown in Fig.~\ref{fig6} (b). The results demonstrate that the average emission sequence of neutrons and protons is almost the same and is insensitive to the centrality. In Fig.~\ref{fig6} (c), the momentum correlation function for $p$-$\bar{p}$ pair with $v_{p}$ $>$ $v_{\bar{p}}$ is slightly higher than one with the reverse situation. The slight asymmetry of velocity difference ($\Delta$v) spectra for $p$-$\bar{p}$ pairs is shown in Fig.~\ref{fig6} (d), which  indicates that the mean order of emission sequence between proton and antiproton may be a little different but  is not sensitive to the centrality. In Fig.~\ref{fig7} (a), the momentum correlation functions for $n$-$p$ pairs with $v_{n}$ $>$ $v_{p}$ are always similar to one with the reverse situation with increasing system-size. The symmetry of velocity difference ($\Delta$v) spectra for $n$-$p$ pairs in different systems is shown in Fig.~\ref{fig7} (b). The comparison of velocity-gated momentum correlation functions illustrates that the average emission sequence between neutrons and protons is always identical for  different centrality and system-size, which 
is also learned from  their ratios in Fig.~\ref{fig8} (a) and (b). In Fig.~\ref{fig7} (c) and (d), the comparison of velocity-gated momentum correlation functions for $p$-$\bar{p}$ indicates that the mean order of emission sequence between protons and antiprotons may be a little different but has no dependence of system-size, which is also learned by their ratios in Fig.~\ref{fig8} (c) and (d). 

Fig.~\ref{fig9} and Fig.~\ref{fig10}  show centrality and system-size dependences of velocity-gated momentum correlation functions and velocity difference ($\Delta$v) spectra of $p$-$d$, $p$-$t$ and $d$-$t$ pairs, respectively. For $p$-$d$ and $p$-$t$ pairs, the momentum correlation functions with $v_{p}$ $<$ $v_{d}$ ($v_{p}$ $<$ $v_{t}$) are stronger than the ones with the reverse situation $v_{p}$ $>$ $v_{d}$ ($v_{p}$ $>$ $v_{t}$) in Fig.~\ref{fig9}. The comparison of  two velocity-gated correlation strengths gives that the mean order of emission of protons  are  emitted averagely earlier than deuterons and tritons according to the above criteria. The similar trend for $d$-$t$ pairs is not so obvious overall, except for in peripheral collision the momentum correlation function with $v_{d}$ $<$ $v_{t}$ is stronger and deuterons are emitted averagely earlier than tritons. In contrast with  the emission order as shown in many previous results of the  intermediate energy heavy-ion collisions~\cite{Gelderloos1995,RGhetti2003,wtt2019,YijieWang2022}, the average emission sequence of protons, deuterons, and tritons is opposite  for 39 GeV heavy-ion collisions.  Meanwile, Fig.~\ref{fig9} presents velocity difference spectra for $p$-$d$, $p$-$t$ and $d$-$t$ pairs, respectively. The velocity difference spectra  are all asymmetric due to the mean emission order. In addition, an enhanced difference between the momentum correlation functions for $p$-$d$ ($p$-$t$ or $d$-$t$) pairs with $v_{p} > v_{d}$ ($v_{p} > v_{t}$ or $v_{d} > v_{t}$) and ones on the reverse situation at larger centrality, which manifests the larger interval of the mean emission order for unlike light nuclei in peripheral collisions. Their ratios in Fig.~\ref{fig11} (a), (c) and (e) can also illustrate the above phenomenon. The system-size dependence for $p$-$d$, $p$-$t$ and $d$-$t$ pairs can be found by the fact that momentum correlation functions with $v_{p} < v_{d}$ ($v_{p} < v_{t}$ or $v_{d} < v_{t}$) are stronger than the ones with the reverse situation $v_{p} > v_{d}$ ($v_{p} > v_{t}$ or $v_{d} > v_{t}$) in Fig.~\ref{fig10}. Correspondingly, the velocity difference spectra for $p$-$d$, $p$-$t$ and $d$-$t$ pairs are all asymmetric about $\Delta$v = 0 caused by the average emission order in Fig.~\ref{fig10}. Therefore, protons are emitted averagely earliest and deuterons are emitted averagely earlier than tritons in smaller system-size collision. The system-size dependence of the velocity-gated momentum correlation functions is also clearly seen by their ratios in Fig.~\ref{fig11}. 
With decreasing system-size, we can also observe an enhanced difference between the momentum correlation functions for $p$-$d$ ($p$-$t$ or $d$-$t$) pair with $v_{p} > v_{d}$ ($v_{p} > v_{t}$ or $v_{d} > v_{t}$) and the ones with the reverse situation in Fig.~\ref{fig11} (b), (d) and (f). 

\section{SUMMARY}

In summary, with the AMPT model complemented by the Lednick$\acute{y}$ and Lyuboshitz analytical method, we have constructed and analyzed the momentum correlation functions of light (anti)nuclei formed by the coalescence mechanism of (anti)nucleons for heavy-ion collisions with different system sizes and centralities at  $\sqrt{s_{NN}}$ = 39 GeV. We present a comparison of proton$-$proton and proton$-$antiproton momentum correlation functions with the experimental data from the RHIC-STAR collaboration~\cite{Zbroszczyk2019,Siejka2019}. Taking the same transverse momentum and rapidity phase space coverage corresponding to the experimental situation as well as the maximum hadronic rescattering time of 700 $fm/c$  in AMPT, it is found that the $p$-$p$ and $p$-$\bar{p}$ momentum correlation functions simulated by the present model can match the experimental data. We further study centrality and system-size dependence of momentum correlation functions for identical and nonidentical light (anti)nuclei pairs, respectively, which is in the condition of the maximum hadronic rescattering time of 100 $fm/c$  in AMPT. The shape of momentum correlation functions for light (anti)nuclei pairs is consistent with previous works~\cite{Star-nature,wtt2018,wtt2019,Zbroszczyk2019,Siejka2019,YijieWang2022}, which is caused by both QS and FSI. The similar structure between light nuclei momentum correlation functions and anti-ones indicates that the interaction between them are the same, which has been confirmed in Ref.~\cite{Star-nature} only about proton and antiproton. The centrality dependence of momentum correlation functions for light (anti)nuclei is investigated by $_{79}^{197}\textrm{Au}+_{79}^{197}\textrm{Au}$ collisions at different five centralities of $0-10$ $\%$, $10-20$ $\%$, $20-40$ $\%$, $40-60$ $\%$, and $60-80$ $\%$ at $\sqrt{s_{NN}}$ = 39 GeV. It is found that with increasing centralities from center to periphery, the momentum correlation functions for light (anti)nuclei become stronger, which are probably emitted from smaller source. The momentum correlation functions of light (anti)nuclei are  sensitive to system-size through studying $_{5}^{10}\textrm{B}+_{5}^{10}\textrm{B}$, $_{8}^{16}\textrm{O}+_{8}^{16}\textrm{O}$, $_{20}^{40}\textrm{Ca}+_{20}^{40}\textrm{Ca}$ and $_{79}^{197}\textrm{Au}+_{79}^{197}\textrm{Au}$ in central collisions, 
and~used to obtain the emission source-size of light (anti)nuclei which is self-consistent with their system-size.  Momentum correlation functions between nonidentical light nuclei can provide important information about the average emission sequence of them. The average emission time scale between neutrons and protons is almost identical. However, heavier light clusters (deuterons or tritons) are emitted later than protons in the small relative momentum region.  
In the future we can explore further the energy dependence of the average emission sequence of light nuclei and understand the physical interpretation.

\section*{Acknowledgments}
T. T. Wang thanks for discussion with Ms. Yi-Ling Cheng for the AMPT data. This work was supported in part by the National Natural Science Foundation of China under contract Nos.  11890710, 11890714, 11875066, 11925502, 11961141003, 11935001, 12147101 and 12047514,  the Strategic Priority Research Program of CAS under Grant No. XDB34000000, National Key R\&D Program of China under Grant No. 2018YFE0104600 and 2016YFE0100900,  Guangdong Major Project of Basic and Applied Basic Research No. 2020B0301030008, and the China PostDoctoral Science Foundation under Grant No. 2020M681140.

\end{CJK*}
\end{document}